\documentclass[11pt]{article}
\usepackage{epsf,psfig,epsfig,amsmath}

\newcommand{\hcenterpage}[1]{
 \setlength{\textwidth}{#1}
 \setlength{\evensidemargin}{3.25in}
 \addtolength{\evensidemargin}{-0.5\textwidth}
 \setlength{\oddsidemargin}{\evensidemargin}
 \setlength{\marginparwidth}{1.74cm}
 \addtolength{\marginparwidth}{-\marginparsep}
 \addtolength{\marginparwidth}{\evensidemargin}
}

\newcommand{\vcenterpage}[1]{
 \setlength{\textheight}{#1}
 \setlength{\topmargin}{4.3in} 
 \addtolength{\topmargin}{-0.5\textheight}
 \setlength{\topskip}{0cm}
}

\renewcommand{\baselinestretch}{1.1}

\hcenterpage{6.75in}
\vcenterpage{9.25in}
\title{Simulations of the adiabatic quantum optimization for the Set Partition Problem.}

\author{V.N. Smelyanskiy$^*$, U. V. Toussaint, D.A. Timucin \\
Computational Science Division\\ NASA Ames Research Center\\
$^*\,$ vadim@quantum.arc.nasa.gov}

\begin{document}
\maketitle
\date{}

\begin{abstract}
We  analyze the  complexity of the quantum optimization
algorithm based on adiabatic evolution for the NP-complete set partition
problem. We introduce a  cost function defined on a logarithmic
scale of the partition residues so that the  total number of
values of the cost function is of the order of the problem size.
We simulate the behavior of the  algorithm by numerical
solution of the time-dependent Schr{\" o}dinger equation as well
as the stationary equation for the adiabatic eigenvalues. The
numerical results for the time-dependent quantum evolution
indicate that the complexity of the algorithm scales exponentially
with the problem size. This result appears to contradict the recent
numerical results for complexity of quantum adiabatic algorithm applied to
a different NP-complete problem (Farhi et al, {\it Science} 292, p.472 (2001)).
\end{abstract}

\renewcommand{\baselinestretch}{1.5}
\normalsize

\thispagestyle{empty}
\section{Introduction}
\label{sec:Intro}

 Most common computationally intensive tasks encountered in
practice may be formulated as combinatorial optimization problems
(COPs), many of which are found to belong to the algorithmic class
\emph{nondeterministic-polynomial complete} (NP-complete)
\cite{Garey}. The NP-complete problems are computationally hard -
they are characterized (in the worst cases) by exponential scaling
of the running time or memory requirements with the problem size.
A special property of the class is that any NP-complete problem
can be converted into any other NP-complete problem in polynomial time on
a classical computer; therefore, it is sufficient to find a
deterministic algorithm that can be guaranteed to solve all instances
of just one of the NP-complete problems within a polynomial time
bound.

An instance of a COP of size $n$ may be encoded using bit
strings $\textsf{z} = z_0 \, z_1 \, \cdots \, z_{n-1}$,   $\,\,z_j
= 0, 1$, with a corresponding value of the cost function (or
``energy'') $E = E_{\textsf{z}}$ for each string.  The objective
is to find the bit string(s) with the minimum cost (and the
corresponding cost value). In quantum computation, bits $z_j$ are
replaced by spin-$\frac{1}{2}$ \emph{qubits}; the qubit states
$|0\rangle$ and $|1\rangle$ are eigenstates of the $\pm z$
component of the $j$-th spin, respectively.  The Hilbert space of
a quantum \emph{register} with $n$ qubits is spanned by $N = 2^n$
basis vectors $|\textsf{z}\rangle = |z_0\rangle \otimes \cdots
\otimes |z_{n-1}\rangle$.

\section{Optimization by Adiabatic Quantum Evolution}
\label{sec:opt}

Following \cite{Farhi,Farhiapp,Cerf}, we consider a quantum evolution of duration $T$ based on the time-dependent
Hamiltonian
\begin{equation}
H(t) = \alpha(s)\, V + \beta(s)\, H_{P}, \quad s\equiv s(t). \label{farhi}
\end{equation}
\noindent
Here, $H_P = \sum_{\textsf{z}} E_{\textsf{z}}
|\textsf{z}\rangle \langle \textsf{z}|$ is the ``problem''
Hamiltonian that embodies the problem structure in its energy
spectrum and eigenstates, the summation being performed over all $N$ $n$-bit
strings, and $V$ is a ``driver'' Hamiltonian that is constructed
in such a way as to cause transitions between those states -
essentially an Ising-type spin Hamiltonian corresponding to 1- and
2-gate operations:
\begin{equation}
V = -\sum_{i=0}^{n-1} B_i \sigma_{x}^{i} - \frac{1}{2} \sum_{i, \,
j=0}^{n-1} J_{ij} \sigma_{x}^{i} \sigma_{x}^{j}.  \label{driver}
\end{equation}

Coefficients $\alpha\left(s\left(t\right)\right)$ and $\beta(s(t))$ vary in time in such
a way that at the initial instant of time $H(0)=V$ and at the
final instant $H(T)=H_P$. A particular choice of the coefficients
is \cite{Farhi}
\begin{equation}
\alpha(s)=1-s, \quad \beta(s)=s, \quad s(t)=t/T\label{alpha}
\end{equation}
\noindent The total Hamiltonian (\ref{farhi})  produces a
nontrivial quantum evolution from some initial (superposition)
state $\psi(0)$ to a final (solution) state $\psi(T)$.  If no
knowledge about the solution is available \emph{a priori}, then
the initial state may be chosen as the symmetric state  (cf.
\cite{Grover,Farhi})
\begin{equation}
\psi(0) =
2^{-n/2} \sum_{\textsf{z}=0}^{2^{n}-1} |\textsf{z}\rangle. \label{ini}
\end{equation}\noindent
This choice is appropriate provided (\ref{ini}) is a ground state
of $V$ (e.g.,   $B_i, J_{ij} \geq 0$).
Now, if $T$ is sufficiently large, then functions $\alpha(t/T)$ and $\beta(t/T)$
vary in time slowly and the system will remain in the
instantaneous  (adiabatic) ground state of $H(t)$ during its
entire evolution $0 < t < T$ (cf.  \cite{Farhi}). Accordingly,
$\psi(T)$ will be a superposition of states $|z\rangle$ corresponding to
the ground state of $H_P$. It is clear that in this case a measurement
performed on the quantum register at $t = T$ will find with
certainty one of the solutions of COP. In this case the complexity
of the quantum algorithm is determined by its duration $T$. If we
expand the wavefunction of the system $\psi(t)$ in the basis of
the adiabatic eigenfunctions $\Psi_k(t)$ of the Hamiltonian $H(t)$
\begin{eqnarray}
&&H(t) \Psi_k(t) = g_k(t) \Psi_k(t), \quad k=0,1,\ldots,2^{n}-1.\label{adiab} \\
&&\psi(t)=\sum_{k=0}^{2^n-1}\,  C_k(t) \Psi_k(t) \exp\left(-i/\hbar \int^{t}_{0} dt' g_k(t')\right),
\label{corr} \\
\end{eqnarray}
\noindent
then adiabatic approximation corresponds to $\psi(t) \propto \Psi_0(t)$ (up to the oscillating
phase factor). Coefficients $C_k(t)$ with $k>0$ correspond to nonadiabatic corrections.
Using perturbation theory in the basis of eigenfunctions  $\Psi_k(t)$ the total probability $p_{n-ad}(t)$
of {\itshape not\/} finding the system at the instant $t$ in its adiabatic ground state equals
\begin{eqnarray}
&& p_{n-ad}(t)=\left(\frac{\dot {s}(t)}{s(t)}\right)^2\,
\sum_{k=1}^{2^n-1}\, \frac{|V_{0k}|^{2}}{[g_k(t)-g_0(t)]^4}
\label{nonad} \\
&& V_{0k}=\langle \Psi_0|V|\Psi_k\rangle.\nonumber
\end{eqnarray}
\noindent Here we used the explicit form of coefficients
$\alpha,\,\beta$ given in (\ref{alpha}). It is seen that during
the quantum evolution coefficients  $C_k(t)\sim \dot s/s \sim 1/T$
and the largest admixture of the exited states into the total
superposition   occurs at the instant  of time when one of
the exited levels $g_k(t)$ closely approaches the ground state
(avoided-crossing). From here the overall criterion for the
adiabatic evolution can be expressed in the well-known form
\begin{equation}
\eta = {\tilde{V}\over T \Delta g_{min}^2} \ll 1, \label{criterion}
\end{equation}
\noindent where $\Delta g_{min}$ is the closest approach of the
ground state to one of the excited states during the evolution - a
minimum gap- and $\tilde V$ is the characteristic energy scale for
the matrix elements of $V$. We note that although instantaneous
nonadiabatic corrections (\ref{nonad}) are quadratic in the
parameter $\eta$ (near the avoided crossing) the probability of
nonadiabatic transitions $W_{n-ad}$ away from the ground state is
exponentially small  in $\eta$ \cite{Landau}. This probability is
defined on an infinite time axis and its logarithm is proportional
to the imaginary part of the integral along the contour in the
complex plane of $t$  that begins and ends on the real time axis and
loops around the complex branching point $t^*$
\begin{equation}
W_{n-ad}~\exp\left(-|{\rm Im} \oint\, g_0(t) dt|\right).
\label{imagin}
\end{equation}
\noindent Here $t^*$ correspond to one of the roots of the
equation
\begin{equation}
g_k(t^*)=g_0(t^*)\label{roots}
\end{equation}
\noindent that provides the smallest value for the exponential in
(\ref{imagin}) (out of all possible complex solutions of
(\ref{imagin}) for different excited states $\Psi_k$). In a
standard (Landau-Zener) theory of nonadiabatic transitions the value
of the exponent is approximately of the order of the parameter
$\eta$ in (\ref{criterion}), and therefore it is the size of the
minimum gap $\Delta g_{min}$ that determines the condition for {\itshape T\/}
and hence the complexity of the quantum adiabatic search algorithm
according to \cite{Farhi}. We note finally that, as pointed out in
\cite{Cerf}, the improved complexity of the adiabatic algorithm is
determined by the instantaneous rate $\dot s(t)/s(t)$ of the variation
of the control parameter $s(t)$ near the avoided crossing. We will
not discuss in this paper such modifications and focus primarily on
intrinsic properties of the quantum system in question.

\section{Set Partition Problem}
\label{sec:spp}

In this paper, we will analyze the complexity of the adiabatic quantum
optimization for the  \emph{set partition problem}
(SPP), which is one of the basic NP-complete problems of
theoretical computer science \cite{Garey}.  The optimization
version of SPP is to partition a set of $n$ positive integers
$\{\alpha_0, a_1, \ldots, \alpha_{n-1}\}$ into two disjoint
subsets ${\mathcal A}_1$ and ${\mathcal A}_2$ such that the
``residue'' $|\sum_{\alpha_j \in {\mathcal A}_1} \alpha_j -
\sum_{\alpha_j \in {\mathcal A}_2} \alpha_j|$ is minimized. The
complexity of the problem substantially depends on the size of the
integers $\alpha_j$ (see below). It is often customary for the
analysis of the random instances of the problem to introduce 
finite-precision rational numbers $a_j$ that are independently and
identically distributed (i.i.d.) in the unit interval $(0,1]$.
\begin{equation}
\alpha_{j} \leq 2^{b} \,\,\,\,\forall j, \quad  a_{j} = 2^{-b}
\alpha_{j} \in (0, 1]
\end{equation}
\noindent Here $b$ is the total number of bits used to represent the
numbers $a_j$. The values of the partition can be encoded in
binaries by attaching ``sign'' bits $s_j$ to the numbers $a_j$. The
partition residue can be defined as $|\Omega_{\textsf{z}}|$ where
\begin{equation}
\Omega_{\textsf{z}} = \sum_{j=0}^{n-1} s_j a_j,\quad  s_j = 1 - 2
z_j = \pm 1 \quad (z_j =0,1)\label{omega}
\end{equation}
\noindent Here $\Omega_{\textsf{z}}$ is a {\em signed} partition
residue. We note that by definition the problem is symmetric: two
bit strings that can be obtained from each other by flipping  all
the bits ($s_j \rightarrow -s_j$) correspond to two values of
$\Omega_{\textsf{z}}$ that differ only in sign. We note that the
minimum-residue partition(s) may be thought of as the ground
state(s) of the following spin Hamiltonian \cite{Mertens}
\begin{equation}
\Omega_{\textsf{z}}^{2} = \sum_{i, \, j=0}^{n-1} a_i a_j s_i s_j.
\label{cost0}
\end{equation}
\noindent This is an infinite range Ising spin glass with Mattis
type antiferromagnetic coupling, $J_{ij}=-a_i\, a_j$. Infinite
range coupling clearly represents a major  problem with direct
(`analog') physical implementation of this Hamiltonian on a quantum
computer. Therefore one can consider using an oracle-type cost
function $E(\textsf{z})=|\Omega_{\textsf{z}}|$ to implement the
problem Hamiltonian in (\ref{farhi}) for SPP.  The corresponding
unitary transformation will multiply the basis states $|z\rangle$
by phase factors $\exp\left(-i\Delta t \,E(\textsf{z})\right)$
during the elementary discrete steps of the  `continuous-time'
adabatic quantum optimization (\ref{farhi}). Although this approach
is natural for the satisfiability problem \cite{Farhi} it has a
serious limitation for  SPP (as well as some other NP-complete
problems like integer programming, where the precision of
integers is of central importance). To demonstrate this point 
we need to consider the density of states of the partition residues.

\subsubsection{Density of states}

We define the density of states for a given instance of SPP as
follows
\begin{equation}
\rho(\Omega) = \sum_{\textsf{z}} \delta(\Omega -
\Omega_{\textsf{z}}).\label{density}
\end{equation}
\noindent The exact form of $\rho(\Omega)$ depends on a given
instance of SPP (i.e., a particular set of numbers $a_j$). However
we introduce a {\em coarse-grained} density of states
\begin{equation}
{\bar \rho}(\Omega) = \frac{1}{\Delta \Omega}
\int_{\Omega}^{\Omega + \Delta \Omega} \rho (\zeta) \mathrm{d}
\zeta \label{cg}
\end{equation}
\noindent where averaging is over an interval of $\Delta \Omega$
whose size will be determined below.  Using (\ref{omega}) and
(\ref{density}) we can rewrite this expression in the form
\begin{equation}
{\bar \rho}(\Omega) = \frac{2^n}{2 \pi} \int_{-\infty}^{\infty}
{e^{i w (\Omega+\Delta\Omega)}-e^{i w \Omega} \over i w}\,I(w)\,
\mathrm{d} w, \quad\quad  I(w)= \prod_{j=0}^{n-1} \cos(a_j w).
\label{cos}
\end{equation}
 \noindent
Note that  $I(\pi k 2^b)=(-1)^k,\,k=0,\pm1,\ldots$ and $I(w)$
has very sharp maxima (minima) at those points. In their
vicinities the integral in (\ref{cos}) can be evaluated by
steepest descent method for any given problem instance. The sum over
the contributions from different saddle points was obtained by
Mertens \cite{Mertens} in his derivation of the partition function
for the corresponding spin glass model. We emphasize however that
$I(w)$ can have multiple sharp resonances at the intermediate
points $|w|<2^{-b}$. The positions of these resonances are at the
multiples of $\pi/q$ where $q$ is an  {\it approximate} greatest
common divisor (g.c.d.) of
the set of $n$ numbers $a_j$ such that $a_j = f_j q+ r_j$ where
$f_j$ are integers and $r_j$ are residues of the division. Provided
that most of the residues are sufficiently small
\[
{\pi^{2}\over 2} \sum_{j=1}^{p} r_{j}^{2} \leq 1, \quad p \sim n,
\]
the function $I(w)$ will have steep peaks at those points. It can
be shown that in the general case the value of the approximate g.c.d.
for a set of $n$ $b-bit$ numbers inside the unit interval
scales as $2^{-n}$ for $n<b$. Obviously it equals  $2^{-b}$ for
$n>b$. In what follows we will be interested in the
high-precision case $n<b$. We choose the size of the averaging
window $\Delta \Omega \gg 2^{-n}$ and this introduces a cut-off
in the integral (\ref{cost0}) at
\[|w| \ll \pi/\Delta\Omega \ll 2^{n}\]
It follows from above that in this case the values of the g.c.d. will
lie outside the cutoff and corresponding resonances will not
contribute to the integral. The value of the integral can be
estimated near the single remaining maximum at $w=0$. The width of
the maximum near that point is ~ $\delta w ~ n^{-1/2} \ll 1$  and
therefore the window function in (\ref{omega}) works as a step
function in that region. Finally we obtain
\begin{equation}
{\bar \rho}(\Omega) = {2^n \over \sqrt{2 \pi n \sigma^2}} \exp
\left(-{\Omega^2 \over 2 n \sigma^2} \right),\qquad
\sigma^2 = \frac{1}{n} \sum_{j=0}^{n-1} a_j^2.
 \label{density1}
\end{equation}
\noindent Here the variance $\sigma$ is
a ``self-averaging'' quantity, and the coarse-graining is
performed over an interval much larger than the characteristic
separation between neighboring partition residue values
\begin{equation}
\Delta E\, \sim  \, \sqrt{n}\,\, 2^{-n} \end{equation} \noindent
yet much smaller than the scale of variation of ${\rho}(\Omega)$:
$\Delta E
 \ll \Delta \Omega \ll \sqrt{n}$.  We note that in the
high-precision regime ($n < b$), partition residues
$\Omega_{\textsf{z}}$ are irregularly spaced and well separated
from each other (on the scale of $2^{-b}$). However this structure
is being averaged out in (\ref{density1}) and the result indicates
that, in general, no more structure exists on a scale
$\gg 2^{-n}$ other than that given by the Gaussian distribution in
(\ref{density1}).  We note that this distribution is usually
obtained for the SPP using averaging over different instances of
the problem  (cf. \cite{Fu,Ferreira}, \cite{Mertens}(\emph{b}));
here we recovered it as a coarse-grained distribution for a {\em
given} instance which is more consistent with our goal of studying
the complexity of the adiabatic quantum optimization algorithm \cite{note2}.

\subsection{Cost function}
Computational complexity of SPP depends critically on the number
of bits $b$: numerical simulations with independent identically
distributed (i.i.d.) random $b$-bit numbers $a_j$ show
\cite{Walsh,Korf} that the solution time grows  exponentially with
$n$ for $2^{n - b} \ll 1$ (high-precision, computationally `hard
phase'), and polynomially for $2^{n-b} \gg 1$ (low-precision,
computationally `easy phase'), exhibiting a behavior similar to a
phase transition \cite{Mertens}(\emph{a}).

In the low-precision phase, values of $\Omega_{\textsf{z}}$ are
equally spaced (in $2^{-b}$) and strongly degenerate each
corresponding to (roughly) $2^{b-n} \gg 1$ number of
bit-strings. This degeneracy grows exponentially with n if $b$
remains fixed. The total number of solutions with zero residues
accumulate correspondingly and this is why the complexity
eventually becomes polynomial in $n$. The quantum algorithm suggested
in \cite{Raedt} directly computes the density of states (\ref{density}) of the SPP
and is efficient in finding the number of
solutions in the low-precision case. In this case it is also
feasible to use a cost function
$E(\textsf{z})=|\Omega_{\textsf{z}}|$ (provided the number of 
possible values does not grow exponentially with $n$).

The situation is qualitatively different in the high-precision
case. Implementation of the approach based on the above cost
function will require a  quantum computer  using exponentially high
precision  physical parameters (external fields, etc) to control
small differences {\em in the phases} of unitary transformations on
the scale at least $~2^{-n}$. This is a technical difference from
the constraint satisfaction problem in which the cost function
generally takes only the set of values that scales polynomially
with $n$; the size of the set equals  the total number of
constraints $m$ (the computationally most difficult case
corresponds to $m \sim n$ and the case of $m \sim 2^{n}$ is not of
general interest there). To avoid the above restriction in
the implementation of the adabatic quantum optimization algorithm for SPP, we
introduce a cost function $E(\Omega)$ based  on a logarithmic
scale of the partition residue values:
\begin{eqnarray}
&&E(\Omega) = 0 , \quad {\rm for} \quad 0\leq |\Omega|/\Delta < 1,
\label{cost2}\\ && E(\Omega) = k ,  \quad {\rm for}
\quad 2^{k-1}\leq |\Omega|/\Delta < 2^{k}, \quad k=0,1,\ldots
L,\nonumber\\ &&2^{L-1} \leq A/\Delta < 2^{L}, \quad A=\sum_{j=0}^{n-1}
a_{j}.\nonumber
\end{eqnarray}
\noindent Since the density of states is linear at $\Omega \ll
\sqrt{ n}$ number of states $d_k$ per energy level
$E(\Omega_{\textsf z})=k$ will grow exponentially with $k$ in that
range ($\sim 2^k$). The total number of levels $L$ depends on the
value of $\Delta$. Using the density of states (\ref{density1})
for $\Omega_{\textsf{z}}\ll \sqrt n$ one can set
\begin{equation}
\Delta =\sqrt{n} \, 2^{-n} \,K.\label{seed}
\end{equation}
where $K$ is some fixed number (a few dozen) independent of $n$.
The  number of ground states of the problem Hamiltonian
\begin{equation}
H_{P} = \sum_{k=0}^{L-1}E(\Omega_{z})|z\rangle \langle z|
\label{Hp}
\end{equation}
\noindent approximately equals $K$, and the total number of
energy levels $L$ is close to $n$. It can be estimated from (\ref{cost2}) that
$n-L  \sim log_{2}{K}$. The distribution of the
low-lying states $d_k$ with cost function (\ref{cost2}) is
somewhat similar to that in the slightly underconstrained cases of
the satisfiability problem.

\section{Results}
\begin{figure}
\centerline{\psfig{figure=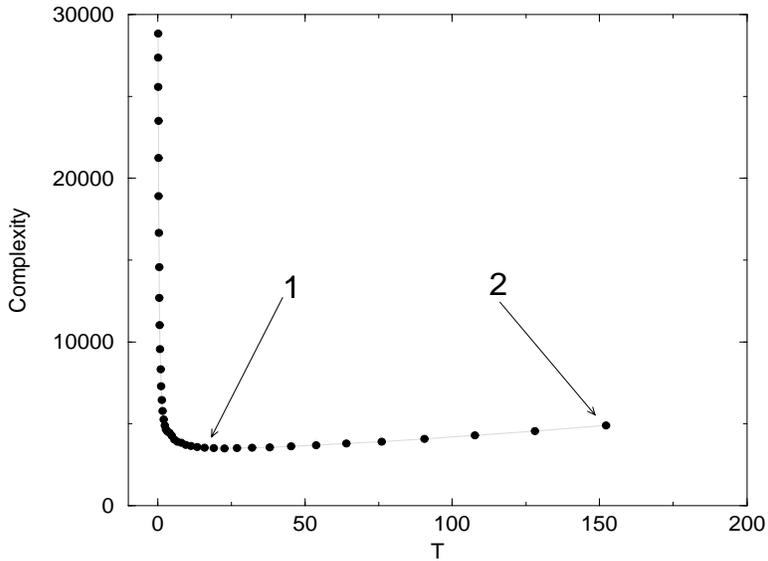,width=4in,height=3in}}
\caption{$C(T)$ {\em vs} $T$ for n=15, precision b=25 bits. Point 1 on the figure corresponds 
to the minimal value of complexity;
the corresponding values are $T^*=22.67$,  $p_0(T^*)=0.15$ and $d_0=22$.  At Point 2 the
total population of the ground level has already reached $p_0(T)=70\%$.
}
\label{fig:complexity15}
\end{figure}

\begin{figure}
\centerline{\psfig{figure=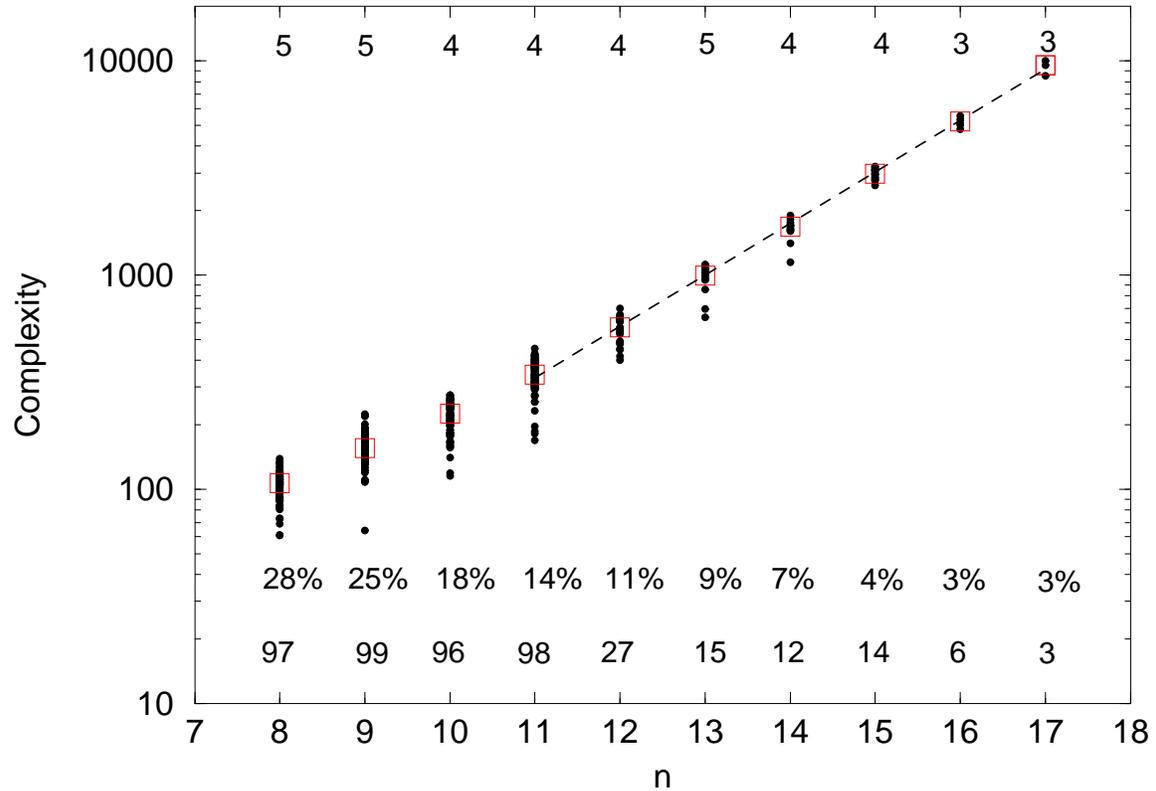,width=6in}}
\caption{$C^{*}$  {\em vs} $n$, precision b=25 bits. Percentage figures correspond to the
total population at the ground level at minimal complexity for a given n
(e.g. 25$\%$ for n=9). Numbers below the vertical sets of points for each $n$ show
the number of trials (e.g., 97 for n=8, 6 for n=16, etc). Numbers
on the top indicate the average values of $d_0$ for all trials at a given $n$.
Median values for the complexity for each $n$ are shown with red squares.
The line is the least squares fit of an exponential function to the median values
between n=11 and n=17.
}
\label{fig:complexity}
\end{figure}

To study the complexity of the adiabatic quantum optimization algorithm
for SPP, we numerically integrate the time dependent
Schr{\" o}dinger equation with the Hamiltonian $H(t)$
(\ref{farhi}), (\ref{driver}) in which we set $B_{i}=1$ and
$J_{ij}=0$. We start from the symmetric initial state (\ref{ini})
and integrate the Schr{\" o}dinger equation in the 
interval $0 \leq t \leq T$. Unlike the approach adopted in
\cite{Farhiapp} we do not set {\em a-priori} a value of success
probability. Instead we introduce a complexity metric for the algorithm
\begin{equation}
{\cal C}(T) = {T+1\over p_0(T)} d_0.\label{complexity}
\end{equation}
\noindent Here $p_0(T)$ is the total probability of finding the system in
its ground level (with $E(\Omega_{\textsf{z}})=0$)  at the end of the algorithm, $t=T$,
and $d_0$ is the number of states at the ground level. The
algorithm has to be repeated on average $d_0/p_0(T)$ number of
times to reach success probability $\approx$ 1.  
A typical plot of
${\cal C}(T)$ for an instance of SPP with $n$=15 numbers is shown in
Fig. \ref{fig:complexity15}. At very small $T$ the wavefunction is
close to the symmetric initial state and the complexity is $\sim 2^{n}$.
The extremely sharp decrease in ${\cal C}(T)$ with $T$ is due to the
buildup of the population $p_0(T)$ in the ground level as quantum
evolution approaches adiabatic limit. At certain $t=T^{*}$ the function
${\cal C}(T)$ goes through the minimum: for $T > T^{*}$ the
decrease in the number of trials $d_0/p_0(T)$ does not compensate
anymore for the overall increase in the runtime  $T$ for each trial.
The minimal  complexity $C^*=C(T^*)$  is
defined via one dimensional minimization over $T$ for a given
problem instance  \cite{adnote}.
\noindent
In Fig.\ref{fig:complexity} we plotted the data for optimal  complexities $C^*$ at different
values of $n$ on a logarithmic scale. Vertical sets of points 
on the plot indicate the results for {\em all} simulation data we currently have
for each $n$. The results indicate that the
median value of complexity $C^*$ scales {\em exponentially} with
$n$; linear fit to the graph gives $\log C^* \approx 0.56\, n$. This
corresponds to the scaling law $C^* \sim 2^{0.8\, n}$.  The
exponential behavior of the algorithm clearly manifests itself for
the larger values of $n\geq$ 11. The scatter in the values of
$\log C^*$ appears to decreases with $n$ however this result is probably due to the
smaller number of data points available  for larger
$n$ values.

\begin{figure}
\centerline{\psfig{figure=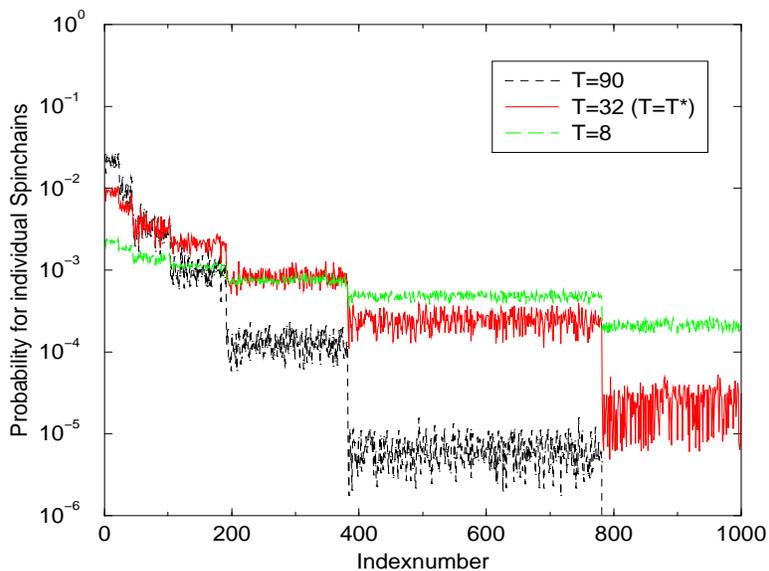,width=4in,height=3in}}
\caption{$|\langle z|\psi(T)|^2$ {\em vs } $z$ for one instance of SPP with n=15.
Note that the values of the index number on the horizontal axis correspond to the positions of 
different bit-strings 
${\textsf{z}}$ 
sorted with  respect to the
partition residue values 
$|\Omega_{\textsf{z}}|$
 (in increasing order). Index number 0 corresponds to the smallest partition
residue. 
 The number of states at the ground
level is $d_0$=22. Curve shown in red  corresponds to the value of $T=T^*=32$ (minimal complexity).
Curve shown in green corresponds to the value of  $T$=8 and black color curve corresponds to $T$=90. 
}
\label{fig:probabilities}
\end{figure}

In Fig. \ref{fig:probabilities} we show the distribution of the
probabilities $|\langle z |\psi(T)|^2$ for different values of $T$
for an instance of SPP with $n$=15 (plots for different $T$ shown with different
colors)and precision b=25 bits. Values of $z$ are ordered with
respect to the corresponding values of the partition residues
$|\Omega_{\textsf z}|$. It is clearly seen on logarithmic scale
that probability distribution forms  'steps' corresponding to
different values of the cost function $E(\Omega_{\textsf z})$ defined
in (\ref{cost2}). Within each step, the distribution of probabilities
does not reveal any structure. The same property holds also for
intermediate times ($t < T$).  Detailed analytical results \cite{VUD}
indicate that it is this absence of structure in
$\psi_z(t)$  that is responsible for the exponential
complexity of the algorithm.
\noindent
\begin{figure}
\centerline{\psfig{figure=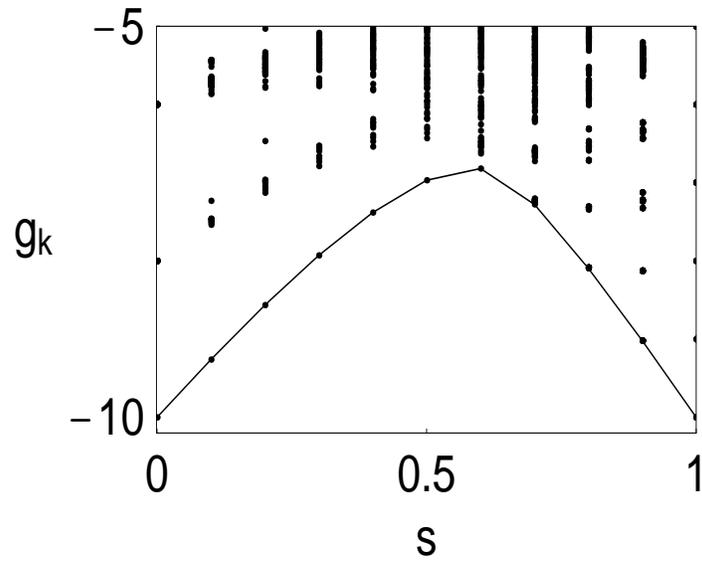, width=4in,height=3in}}
\caption{Adiabatic eigenvalues $g_k$ {\em vs} $s=t/T$ (k=0,1,2,$\ldots$).}
\label{fig:Eigenvalues1}
\end{figure}
\noindent
\begin{figure}
\centerline{\psfig{figure=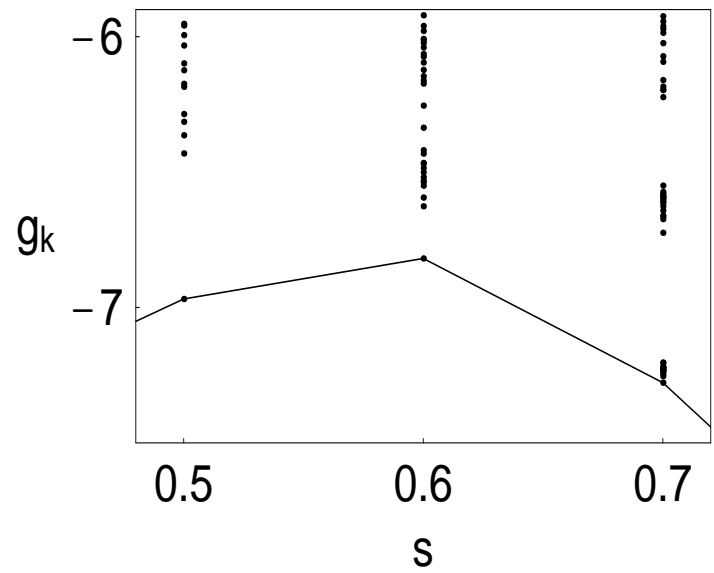,width=4in,height=3in}}
\caption{Magnified version of the Fig. \ref{fig:Eigenvalues1} in the avoided -crossing region.}
\label{fig:Eigenvalues2}
\end{figure}
\noindent
\subsection*{The Stationary Schr{\"o}dinger equation}
In addition to solving the time-dependent Schr{\" o}dinger  equation
we also analyzed the adiabatic solutions of the stationary Schr{\"
o}dinger equation with the same form of the Hamiltonian $H(t)$
(\ref{farhi}) as above. Our preliminary  results were obtained using
Mathematica for modest values of $n \leq$ 10. The results for n=10 
are shown in Figs. \ref{fig:Eigenvalues1} and
\ref{fig:Eigenvalues2}. Figure \ref{fig:Eigenvalues2} represents the magnified
part of Fig. \ref{fig:Eigenvalues1} near the avoided crossing region.
Adiabatic eigenvalues were computed for different
values of the scaled time parameter $s=t/T \in (0,1)$. The solid line
represents the evolution of the ground state eigenvalue between $s=0$ and $s=1$.
The vertical sets of points
correspond to excited adiabatic levels for a given  $s$. At the
beginning ($s=0$) eigenvalues correspond to those of the
Hamiltonian $V$: equally spaced levels -n,-n+2, $\ldots$, n,
corresponding to different number of spin excitations along the $x$
quantization axis. The first excited state is $n-$fold degenerate,
the second is $n(n-1)$, the k-th exited state is $\binom{n}{k}$-fold
degenerate, etc. For $s>0$ the degeneracy is removed. For
$s\rightarrow 1$ the eigenvalue spectrum is the one for the problem
Hamiltonian $H_{P}$ (in \ref{Hp}). We have shifted the energy reference in the Hamiltonian
$H_P$ by $-n$ (cf. also (\ref{farhi})) to match the energy scale for the symmetric case
which emphasizes the avoided crossing region.
In our case the ground state was 13-fold degenerate and
the corresponding eigenvalues merge at $s\rightarrow 1$. The close
approach of these eigenvalues is not relevant for the minimum-gap
analysis since they all end up in the same final level. However
the minimum separation of the instantaneous adiabatic ground state
eigenvalue from the excited state eigenvalues that {\em do not} end up
on the same ground level at $s=1$ is clearly seen in the figures.
Note that the size of this separation is much greater than the
separations between the excited states. This behavior clearly
departs from the standard 2-level avoided crossing picture and is
due to the contributions from the exponential number of terms in
(\ref{nonad}) as will be analyzed elsewhere \cite{VUD}.
We also note that the value n=10 does not correspond to the
exponential scaling regime for the algorithmic complexity that appears to
start for greater $n$  values as follows from the discussion above.

In conclusion, we have performed numerical simulations of the
adiabatic quantum optimization for SPP using a step-like density
of states defined on a logarithmic scale of partition residues.
The results indicate an exponential scaling of the algorithmic
complexity as a function of the problem size. The apparent reason
is the loss of structure in SPP during the effective
coarse-graining over the intervals of partition residues
corresponding to the same cost function values.

%


\begin{thebibliography}{99}
\bibitem{Garey}M.R. Garey and D.S. Johnson, {\em Computers and Intractability. A Guide to the Theory
of NP-Completeness} (W.H. Freeman, New York, 1997)

\bibitem{Farhi}  (\emph{a})  E. Farhi, J. Goldstone, S. Gutmann, and M. Sipser, ``Quantum computation by
adiabatic evolution,'' arXiv:quant-ph/0001106;  (\emph{b}) 
E. Farhi, J. Goldstone, S. Gutmann, J. Lapan, A. Lundgren, and D. Preda,  ``A quantum adiabatic evolution
algorithm applied to random instances of an NP-complete problem'', {\it Science} {\bf 292}, 472 (2001).

\bibitem{Farhiapp}  E. Farhi, J. Goldstone, and S. Gutmann, 
``A numerical study of the performance of a quantum adiabatic evolution algorithm for satisfiability,'' 
arXiv:quant-ph/0007071; A. M. Childs, E. Farhi, J. Goldstone, and
S. Gutmann, ``Finding cliques by quantum adiabatic evolution'',  arXiv:quant-ph/0012104.

\bibitem{Cerf} J. Roland and N. Cerf, "Quantum Search by local adiabatic evolution",
arXiv:quant-ph/0107015

\bibitem{Grover} L. K. Grover, ``Quantum mechanics helps in searching for a needle in a haystack,''
Physical Review Letters {\bf 79}, 325--328 (1997).

\bibitem{Landau} L.D. Landau and E.M. Lifschitz , "Quantum
Mechanics", Pergamon, London (1959).

\bibitem{Mertens} S. Mertens, (\emph{a}) ``Phase transition in the number partitioning problem,''
Physical Review Letters {\bf 81}, 4281--4284 (1998); (\emph{b})
``Random costs in combinatorial optimization,'' {\em Physical
Review Letters} {\bf 84}, 1347--1350 (2000).

\bibitem{Fu} Y. Fu, in {\em Lectures in the Sciences of Complexity}, ed. by D.L. Stein (Addison-Wesley Publishing Company, Reading, Massachusetts, 1989).

\bibitem{Ferreira} F. Ferreira and J. Fontanari, {\em Journal of Physics A} {\bf 31}, 3417 (1998).

\bibitem{note2}
 We also note that occasionally for certain "singular"
instances of the partition problem values of approximate g.c.d.
can be rather large (e.g. $~1/n$), i.e. most of the numbers are
nearly commensurate with each other. In this case intermediate
resonances will be of interest and the density of states will have
an additional structure at low 'frequencies' $\delta w ~ /pi/d$.
We do not consider those instances in a present paper.

\bibitem{Walsh} I.P.Gent and T. Walsh, {\em Comp. Intell.} {\bf
14}, 430 (Blackwell, Cambridge MA, 1998).

\bibitem{Korf} R.E. Korf, {\em Artif. Intell.} {\bf 106}, 181
(1998).

\bibitem{Raedt} H. De Raedt et al,{\em Phys. Lett. A} {\bf 290},5-6, 
p. 227-233 (2001).

\bibitem{adnote} We note that an additional optimization can be
done if one defines a complexity $C$ similar to (\ref{complexity})
but uses an  intermediate time instance $t$ instead of $T$ there.
In this case quantum algorithm is terminated  at 
the instance  $t=t^*(T)\leq T$ for each $T$ when minimal complexity
is reached and then additional minimization over $T$
is performed. Our results indicate that this method does 
provide further improvement for the overall complexity but takes a 
prohibitely long time to perform numerical  optimization of $C$.
 
\bibitem{VUD} V.N. Smelyanskiy, U.V. Toussaint, D.A. Timucin, to
be submitted.
\end{thebibliography}
\end{document}